\documentclass[%
 reprint,
 amsmath,amssymb,
 aps,
pra,
]{revtex4-2}

\usepackage{graphicx}% Include figure files
\usepackage{dcolumn}% Align table columns on decimal point
\usepackage{bm}% bold math
\usepackage{makecell} % for \makecell in tables
\usepackage{tikz}
\usepackage{subcaption}
\usepackage{caption}
\usepackage{amsmath}
\usetikzlibrary{arrows.meta}
 \usetikzlibrary{calc}

\begin{document}

\preprint{APS/123-QED}

\title{Long Term Dynamics and Stability Studies for the High-energy Electron Cooler}% Force line breaks with \\

\author{Jonathan Unger}
 \email{jeu8@cornell.edu}

\author{Georg H. Hoffstaetter}%
 
\affiliation{%
 Cornell University, Ithaca NY , USA
}%

\author{Alexei Fedotov}
\author{Yichao Jing}
\author{Dmitry Kayran}
\author{Jorg Kewisch}
\author{Sergei Seletskiy}
\affiliation{Brookahven National Laboratory, Upton NY, USA}

\date{\today}% It is always \today, today,
             %  but any date may be explicitly specified

\begin{abstract}

The Ring Electron Cooler (REC) is being developed for the Electron-Ion Collider to provide cooling at the top proton energy of 275GeV to reduce emittance growth from intra-beam scattering (IBS) and other effects. Cooling electrons circulate in a 150 MeV storage ring equipped with strong damping wigglers. While these wigglers counteract emittance degradation driven by intra-beam scattering and proton-electron  beam-beam scattering, they introduce significant nonlinear dynamics which can affect electron beam lifetime. In this paper we present the nonlinear optimization of the REC, including optimized focusing wiggler fields, chromatic correction, misalignments and orbit correction and space-charge studies. A novel sextupole-like wiggler field configuration is shown to substantially reduce chromatic aberrations compared with quadrupole-like focusing. Transverse and longitudinal dynamic apertures exceeding 5$\sigma$ are achieved in the presence of realistic alignment and BPM errors. Space-charge simulations for a Gaussian beam using the Bassetti–Erskine model show manageable tune spread and limited equilibrium emittance growth. These results demonstrate the feasibility of REC operation under the required cooling parameters.

\end{abstract}

%\keywords{Suggested keywords}%Use showkeys class option if keyword
                              %display desired
\maketitle

%\tableofcontents
\section{introduction}

In order to achieve high luminosity and beam lifetime, the Electron-Ion Collider (EIC) would strongly benefit from continuous cooling of the proton beam at top energy to compensate emittance growth driven primarily by intra-beam scattering (IBS) \cite{CDR}. 

Cooling intense proton bunches at high energy is an extremely challenging task. While electron cooling is a suitable technique, scaling it to high energy requires an electron beam with low emittance and high average current. Recirculating electrons in a storage ring solves the high-current problem, but gives rise to electron emittance degradation caused by non-linear dynamics and various collective effects. This work addresses these challenges for the proposed high-energy electron cooler which employs electron storage ring.

A ring electron cooler (REC) \cite{osti_2573763} is under development as a candidate cooling system for protons in the EIC hadron storage ring (HSR) at energies at 275 GeV. In electron coolers \cite{electronCooling}, electrons co-propagate with hadrons at the same average velocity in a straight section of the hadron storage ring called the cooling section (CS). As a result, the hadrons experience a friction force that, over multiple revolutions in the ring, reduces the 6D phase-space volume occupied by the hadron bunches.

The REC is an RF-based, non-magnetized electron cooler \cite{osti_1602464,osti_1575964,Fedotov:2022ole,PhysRevAccelBeams.23.021003,PhysRevAccelBeams.23.110101}.
In the REC, an electron beam circulating in a 150 MeV storage ring cools the protons in a 170-m-long cooling section. Compared with single-pass cooling systems, the REC substantially reduces demand on the injector setup by reusing electron bunches over many turns. However, this introduces several challenges associated with maintaining acceptable electron beam quality over the damping time.

The dominant limitations arise from emittance growth in the electron beam due to intra-beam scattering (IBS) and beam-beam scattering (BBS) with the proton beam. Because the electron energy required for cooling 275 GeV protons is only 150 MeV, synchrotron radiation damping in the ring arcs is weak, requiring the use of strong damping wigglers to maintain equilibrium emittance. These wigglers must simultaneously provide substantial radiation damping while minimizing the IBS emittance growth in their field which requires a strong horizontal focusing in the wigglers.

The strong focusing has the potential to introduce strong nonlinear dynamics and chromatic effects that can significantly reduce momentum and transverse dynamic aperture. In addition, the relatively low electron beam energy and high bunch charge produce large direct space-charge tune shifts, while image-current effects in the narrow-gap wigglers will further influence beam dynamics. The combined effects of strong nonlinear optics, collective effects, and maintaining stability with realistic magnet errors present a challenge for the realization of a practical REC lattice.

In this work, we present the nonlinear optimization of the storage ring electron cooler. Particular emphasis is placed on the design of focusing wigglers, dynamic aperture optimization, and the impact of collective effects including direct and indirect space charge through image currents. A novel sextupole-like focusing wiggler field is shown to significantly reduce chromatic effects compared with a simple quadrupole-like focusing approache, enabling substantial improvement in dynamic aperture. Orbit correction and alignment tolerance studies demonstrate that the optimized lattice remains operational under realistic machine errors. Tracking simulations including space charge show that the resulting tune spread and emittance growth remain manageable over multiple damping periods.

These results demonstrate the feasibility of REC operation within the required EIC cooling parameters and establish a framework for further optimization of ring-based electron cooling systems.

\section{Ring Design}

The REC is designed to provide continuous electron cooling for protons in the HSR at 275 GeV. Electron cooling at this energy requires a 150 MeV electron beam co-propagating with the proton beam through a 170 m cooling section. The REC lattice is therefore designed to simultaneously satisfy cooling requirements, radiation damping constraints, nonlinear beam dynamics limits, and the geometric constraints of the existing EIC tunnel.

The ring adopts a compact trapezoidal geometry with a total circumference of 426 m, allowing the full cooling section to overlap the proton beam for 170 m while maintaining the harmonic timing relationship required between the electron and proton bunches. The primary lattice parameters are summarized in Table \ref{base_parameters}. The Ring Layout is shown in Fig. \ref{RECOptics}. The REC optics is shown in Fig. 2.

\begin{table}[htb]
    \caption{Baseline parameters}
    \begin{tabular}{|l c|}
        \hline
       $N_e$  &  1.3$\cdot 10^{11}$\\
        Relativistic $\gamma$ & 293\\
        Ring circumference [m] & 426\\
        Cooling section length [m] & 170\\
        Cooling section $\eta_x$ [cm] & 100\\
        Cooling section $\beta^*_{x,y}$ [m] & 180, 160\\
        Number of wigglers & 18\\
        Wiggler length [m] & 4.2\\
        Wiggler field [T] & 2.4\\
        Wiggler period [cm] & 23\\
        RF frequency [MHz] & 98\\
        RF voltage [kV] & 2.1\\
        Bunch Length [cm] & 13\\
        Momentum spread & $9.8\times10^{-4}$\\
        Emittance (x,y) [nm] & 7.8, 7.8 \\
        Fractional tune (x,y) & 0.16, 0.15 \\
        Space charge tune shift (x,y) & -0.08, -0.12\\
        \hline
      
    \end{tabular}
    
    \label{base_parameters}
\end{table}

\begin{figure}[htb]
    \centering
    \includegraphics[width=1\linewidth]{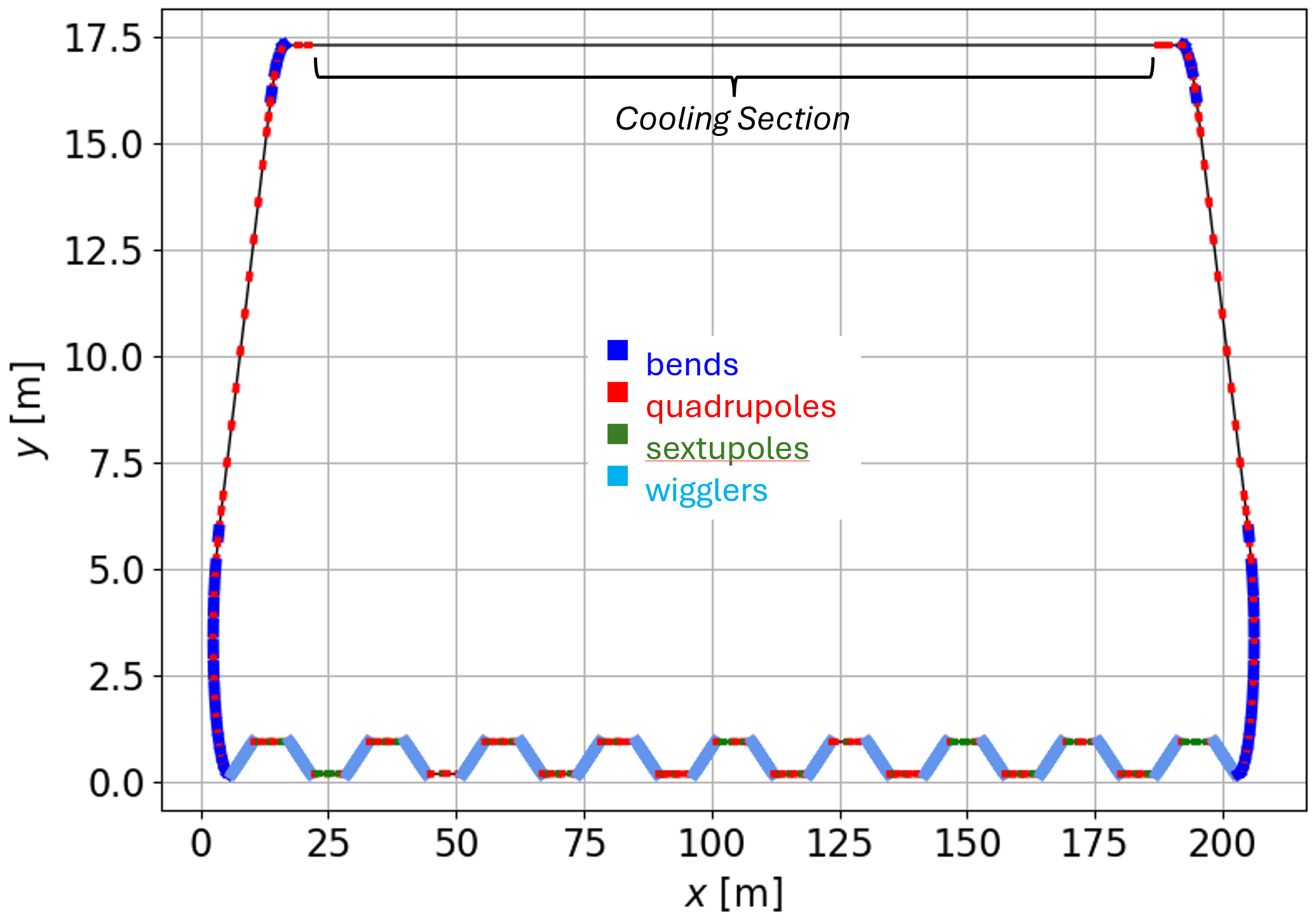}
    \caption{REC floor plan}
    \label{RECFloorPlan}
\end{figure}

\begin{figure}[ht]
    \centering
    \includegraphics[width=1\linewidth]{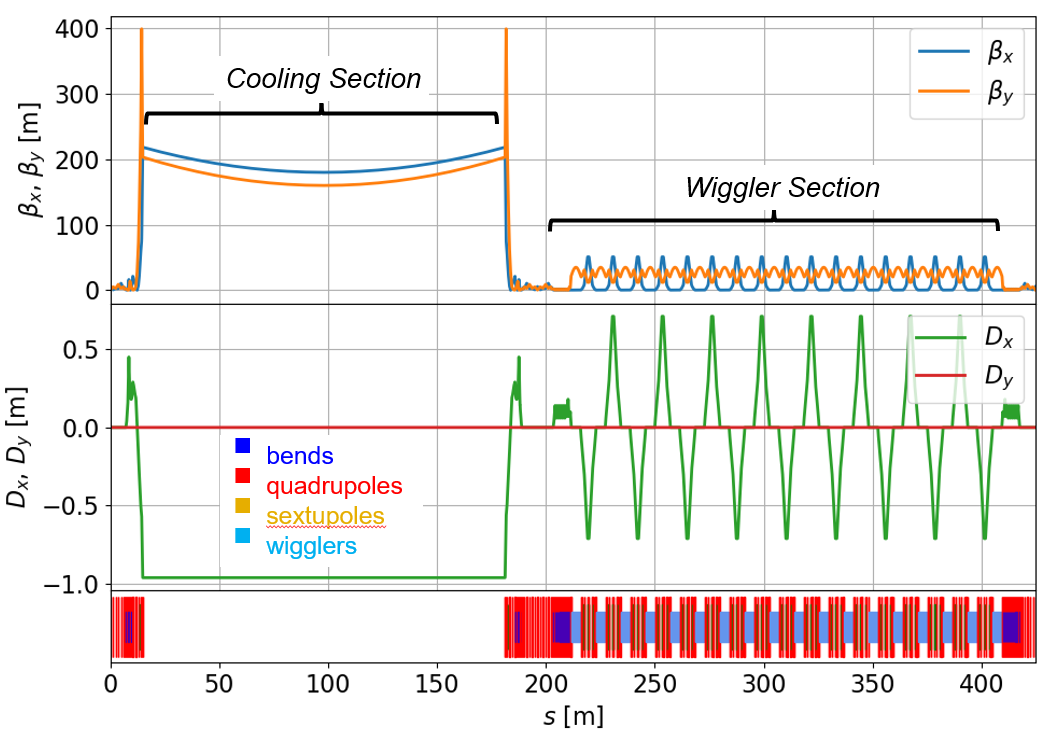}
    \caption{REC optics}
    \label{RECOptics}
\end{figure}

The EIC requires 2 hours of horizontal and 3 hours of longitudinal cooling time for protons at top energy. To achieve the required cooling rates, the REC redistributes the cooling from the longitudinal to the transverse direction. This redistribution requires non-zero ion and electron dispersions in the cooling section \cite{osti_2573763}. The Twiss parameters in the CS are determined by optimizing the system to provide the necessary cooling times while minimizing the electron bunch charge \cite{Seletskiy:2026jnu}. Finally, the equilibrium emittance and energy spread of the electron bunches are dictated by the balance of intra-beam scattering (IBS), beam-beam scattering (BBS), and radiation damping.

Because synchrotron radiation damping at 150 MeV is weak in the ring arcs, radiation damping is generated primarily through 18 wigglers, each 4.2 m long with a peak field of 2.4 T. These wigglers provide the damping required to counteract emittance growth from space charge, IBS, and BBS with the proton beam. In order to reduce IBS-driven emittance growth inside the wigglers, focusing is incorporated into the wiggler fields.

The sections between wigglers contain the sextupoles and octupoles used for chromatic correction and nonlinear optimization. The lattice optics are configured such that the wigglers generate substantial dispersion derivative while maintaining nearly zero dispersion at the insertion boundaries, allowing efficient chromatic correction with moderate sextupole strengths.

The REC operates with a 98 MHz RF system chosen to satisfy the timing constraints of the proton beam while maintaining the target bunch length and longitudinal stability. An additional harmonic cavity configuration has also been considered to reduce peak charge density and mitigate collective effects if needed.

\section{Wiggler Fields}

In order to cool the electron beam and counter the emittance increase caused by the proton bunches and IBS, strong 2.4T wigglers are used for radiation damping. 18 of these wigglers are used, with each being 4.2m long, accounting for almost a fifth of the storage ring circumference. In order to keep IBS contributions small in the wigglers, focusing is desired to minimize the function

\begin{equation}\label{curlyh}
    \mathcal{H}_x=\gamma_x \eta_x^2 + 2 \alpha_x \eta_x \eta_x' + \beta_x \eta_x'^2
\end{equation}

\noindent which will help reduce emittance growth in the wigglers.

By choice of wiggler period and strength, $\eta_x$ can be kept small in the wiggler, reducing the first term. However, $\eta_x'$ remains large, the final term is reduced by adding focusing in the wigglers to keep $\beta_x$ small.

Multiple fields can give wiggling and focusing, but with different non-linear effects. A smart choice of fields can reduce harmful effects, leading to a substantial improvement in DA.

The REC uses high-field wigglers at relatively low energy, making the standard 'short-lens' treatment of the wiggler optics inadequate. In fact, there is a substantial phase advance per REC wiggler, which results in uniquely interesting optical properties for these devices. A detailed derivation of this optics is given in \cite{osti_2428917}. Below, we use the results of \cite{osti_2428917} relevant to the considerations of this paper.

\subsection{Quadrupole-like field}

The simplest choice of fields is a sinusoidal main field with a quadrupole-like focusing term.

\begin{equation}
     B_x=-B_0\cos{\left(k_q x\right)}\sinh{\left(k_q y\right)}
\end{equation}
 
\begin{multline}
    B_y=B_0\cosh{\left(k_z x\right)}\sin{\left( k_z z \right)}\\ - 
    B_0\sin{\left( k_q x\right)}\cosh{\left(k_q y\right)}
\end{multline}
    
\begin{equation}
        B_z = B_0 \sinh{\left(k_z y\right)}\cos{\left(k_z z\right)} 
\end{equation}

An analytic approximation of the chromaticity of wiggler types with different focusing methods has been derived\cite{wigglerChrom}, which closely followed the numerical results from these wigglers. For the wiggler with the superimposed quadrupole-like field, the analytic approximation for the chromaticities is

\begin{equation}
    \xi_{x}=-\frac{\sqrt{B_0k_q}N_p\lambda}{4\pi \sqrt{B\rho}}
\end{equation}
\begin{equation}
    \xi_y=-\frac{B_0N_p\lambda}{2\pi\sqrt{2}B\rho}\frac{1-k_qB\rho/B_0}{\sqrt{1-2k_qB\rho/B_0}}
\end{equation}

For $\xi_y$, it is seen that as $k_q$ approaches $B_0/\left(2B\rho\right)$ there is a singularity, which occurs when $k_q\rightarrow k_z$. In the case of the REC, a choice of $k_q$ near this value is necessary to achieve the required focusing in the wigglers, making this wiggler design undesirable.

\subsection{Sextupole-like Field}

The choice of field arrived at for this lattice is a shaped main field that include a sextupole-like addition:

\begin{gather}\label{wig}
    B_x = \frac{k_x}{k_y} B_0
    \sinh{\left(k_x x\right)}\sinh{\left(k_y y\right)}\sin{\left(k_z z\right)} \\
    B_y = B_0 \cosh{\left(k_x x\right)}\cosh{\left(k_y y\right)}\sin{\left(k_z z\right)} \\
    B_z = \frac{k_z}{k_y} B_0
    \cosh{\left(k_x x\right)}\sinh{\left(k_y y\right)}\cos{\left(k_z z\right)} \\
    k_z^2 = k_x^2+k_y^2
\end{gather}

This achieves focusing as the bunches wiggle through the sextupole-like portion of the field, achieving an amplitude dependent focusing from the main field wiggling that is similar to the focusing amplitude dependent focusing from dispersion in chromatic sextupoles.

Following the same steps as in the quadrupole-like field\cite{wigglerChrom}, an analytic approximation for the chromaticity is:

\begin{equation}
    \xi_x=-\frac{1}{2\pi}\frac{B_0N_p\lambda}{\sqrt{2}B\rho}\frac{k_x}{k_z}
\end{equation}

\begin{equation}
    \xi_y=-\frac{1}{2\pi}\frac{B_0N_p\lambda}{\sqrt{2}B\rho}\frac{\sqrt{k_z^2-k^2_x}}{k_z}
\end{equation}

The singularity in $\xi_y$ is absent for this field configuration, allowing for a significant reduction in the sextupole strengths needed for chromaticity correction leading to substantial improvement of the transverse aperture.

\subsection{Origin of the Chromatic Phase-Advance Singularity}

The pole in the chromatic phase contribution from the wiggler with superimposed quadrupole can be understood by an analysis of the chromatic dependence of each term of the focusing. The vertical equation of motion in this case is

\begin{equation}
    y''=b\cos(k_qx)\sinh(k_qy)-x'b\sinh(k_zy)\cos(kz)
\end{equation}

where $b=B_0/B\rho$ and first and second terms come from the quadrupole and wiggler contributions to the focusing respectively. Linearization of the first term shows the focusing to scale as $K_q\propto b$. The second term, after linearizing in y and averaging over the oscillating contributions of $x’ \cos(k z)$, has a focusing scaling as $K_w\propto x'b$. The primary $x$ wiggling motion scales with $b$, leading to a final scaling as $K_w\propto b^2$. This leads to the equation

\begin{equation}
    y''+(A_1b^2-A_2b)y=0
\end{equation}

\noindent where $A_{1,2}$ are scaling factors that describe the details of the field and averaging. The phase advance accumulated through such a wiggler of length $L$ is then 

\begin{equation}
    \mu_w=L\sqrt{A_1b^2-A_2b}
\end{equation}

The chromatic dependence comes in through $b(\delta)=b_0/(1+\delta)$ giving the leading order chromatic phase advance

\begin{equation}
    \left.\frac{\partial\mu_w}{\partial \delta}\right|_{\delta=0}=-\frac{2A_2b_0^2-A_1b_0}{2\sqrt{A_1b_0-A_2b_0^2}}
\end{equation}

\noindent which becomes singular when $A_1b_0-A_2b_0^2=0$, demonstrating that the chromatic derivative of the phase advance diverges as the net focusing approaches zero.

Strictly speaking, this result describes the chromatic phase advance generated by the wiggler rather than the global machine tune. However, since the machine tune is determined by the accumulated phase advance around the ring, a divergent chromatic derivative of an insertion phase advance produces a correspondingly large contribution to the machine chromaticity whenever the insertion provides a significant fraction of the local focusing. This is the case in the REC where extreme chromaticity was observed in the vicinity of the pole. The singular behavior is therefore a direct consequence of the coexistence of focusing terms proportional to $b$ and $b^2$ and is independent of the specific values of the coefficients $A_1$ and $A_2$, with \cite{osti_2428917} showing the full derivation of the coefficients for these fields.

These approximate formulas were tested against tracking for the strength of wigglers and focusing needed for the REC. A small shift in the location of the singularity in the quadrupole-like focusing was observed, but remained close to the desired wiggler focusing. This shows that wigglers with sextupole-like focusing provide the optics required to suppress horizontal IBS within the devices, while minimizing the wigglers' chromatic contribution. 

While the "strong-field, low-energy" wiggler with a quadrupole-like focusing field is unsuitable for the REC, potential applications for such devices still exist. Such potential applications include creating a mono-chromatic beam for small momentum spread tolerance, momentum dependent resonant extraction, and sensitive tune based diagnostics of momentum spread and bunch momentum. Additional exploration of the chromatic effects of such field configurations could also lead to interesting results in quantities such as the W-function or higher order chromaticity corrections.

\section{Dynamic Aperture optimization}

In order to ensure that the necessary beam lifetime is achieved, the dynamic aperture of the machine must exceed 5$\sigma$ in all dimensions, with excess to allow for decreases from magnet errors, with all tracking being done in full 6D phase space. The DA optimization included variation of sextupoles and octupoles strengths as well as adjustments to the phase advance over a wiggler block.

\begin{figure}[ht]
    \centering
    \includegraphics[width=1\linewidth]{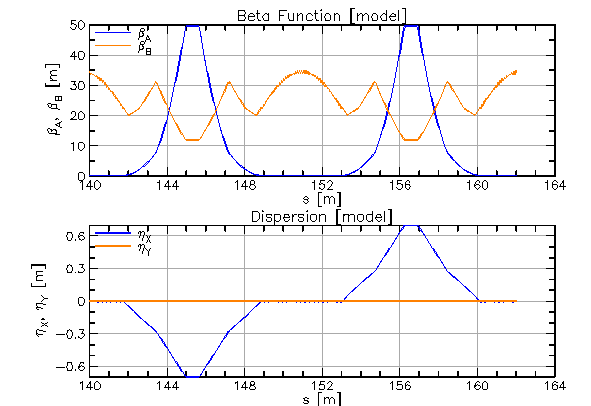}
    \caption{Optics across two wiggler blocks.}
    \label{WigglerSectionOptics}
\end{figure}

In between the wigglers, sextupoles are quadrupoles are placed to facilitate the REC's chromatic correction. Large dispersion at the sextupoles is achieved via choosing the periodic condition for wigglers such that $\eta_x=0$ at both ends. Thus a large $\eta_x'$ coming out of the wigglers is used for the sextupoles, leading the optics of Fig \ref{WigglerSectionOptics}.

The sextupoles in the wiggler section were split into one family per plane. The chromaticity correction is spread mostly among these sextupoles. The large $\beta$ peaks on either side of the cooling section each have an additional sextupole. These sextupoles were used to allow a reduction in the average W-function of the ring, which is defined as the Montague function of expanded optics functions \cite{Montague:443342}

\begin{equation}
    W=\sqrt{\left( \frac{\partial\alpha}{\partial p_z}-\frac{\alpha}{\beta}\frac{\partial \beta}{\partial p_z}\right)^2+\left( \frac{1}{\beta}\frac{\partial \beta}{\partial p_z}\right)^2}
\end{equation}

The sextupoles are located in short straight sections between wigglers. As wigglers and sextupoles are responsible for strong non-linear transverse effects, choosing a proper phase advance across wiggler blocks can enable partial cancellation and improve DA.

\begin{figure}[htb]
    \centering
    \includegraphics[width=1\linewidth]{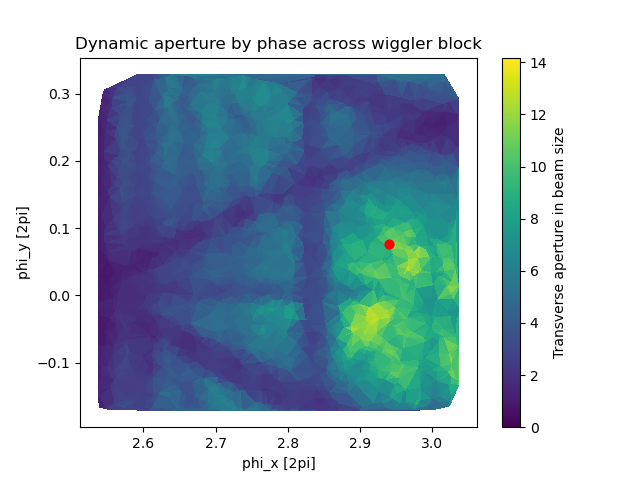}
    \caption{Scan of wiggler block phase advance with color grading showing the transverse aperture and chosen phase advance in red.}
    \label{DAPhase}
\end{figure}

In order to determine the ideal phase advance, phase trombones were inserted into the middle of the wigglers and varied for many DA calculations. Multiple regions of potential phase advances were tried, with the yellow region in Fig. \ref{DAPhase} above $\phi_y=0$ being chosen for good DA and easy matching.

If sextupole optimization did not achieve the required momentum aperture, octupoles were used to reduce the energy-dependent tune spread up to the target momentum aperture. This was done both on an order by order basis with chromaticity as well as reducing spread over the desired range. This effectively controls the tune up to third order.

\begin{figure}[htb]
    \centering
    \includegraphics[width=1\linewidth]{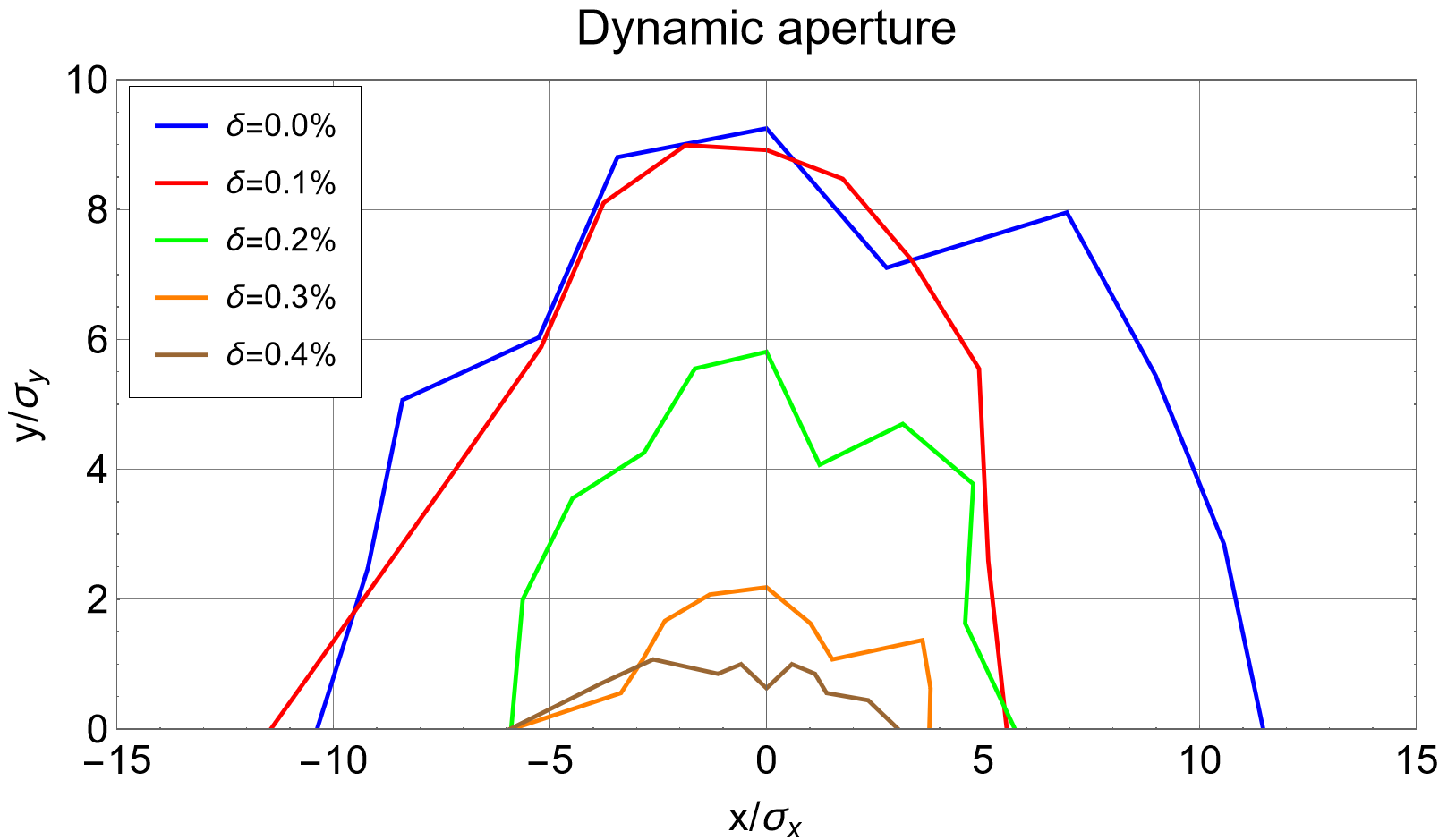}
    \caption{DA after sextupole and octupole optimization}
    \label{RECDAOct}
\end{figure}

After implementing these strategies, The transverse aperture exceeded the target value substantially, while momentum aperture sits just at the target value of $5\sigma_{p_z}$. the transverse size at this momentum aperture is below $1\sigma$, although this can be increased with a minor reduction in max $\delta$ by relaxing the octupole strengths, leading to Fig. \ref{RECDAOct}. Further optimization for momentum aperture is ongoing to provide additional margin for errors.

\begin{figure}[htb]
    \centering
    \includegraphics[width=1\linewidth]{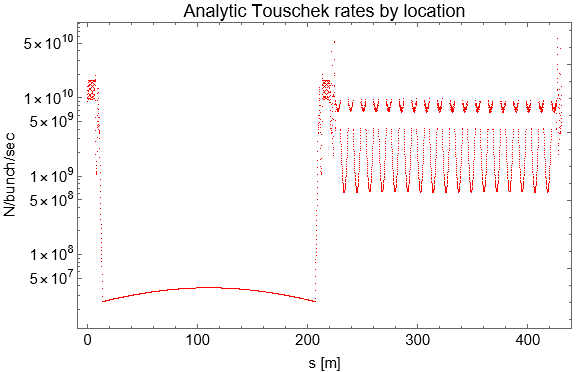}
    \caption{Analytic Touschek rates for the optimized lattice, showing the highest rates in the arcs and wigglers.}
    \label{AnalyticTouschek}
\end{figure}

This momentum aperture leads to a Touschek lifetime of 47sec. With the largest contributions coming from the wiggler section based on the analytic rates in Fig. \ref{AnalyticTouschek}. 

With these verifications, it is seen that the optimizations of the previous section are successful in achieving the goal of transverse aperture, with a small margin of increase still needed for the momentum aperture. 

\section{Sensitivity to misalignments and field errors}

In order to determine the margin provided by this correction scheme, a study of magnet misalignment and field errors was also conducted. First, individual magnet types were tested under each error type separately. This was done in order to determine the relative severity in errors for different magnet types, which would be used to inform the second combined study.

\subsection{Alignment Errors}

Initially, all magnet types were tested for different rms translation errors and rms tilts about the longitudinal axis to meet the criteria of meeting the DA goal in $50\%$ of ran error seeds, with the number of seeds chosen at 1000. No orbit, tune, or other correction was applied at this stage and all combinations of magnet and error type were run individually. Further differentiation between the misalignment of various quadrupoles showed that the quadrupoles near the cooling section are more sensitive than those in the arcs or the wiggler section. These results are used to inform relative tolerances of magnet types when combined with each other and correction and are shown in Table \ref{translationTolerances}.

\begin{table}[htb]
    \centering
     \caption{Tolerances for individual magnet type translations and tilts without orbit correction. Used as baseline for magnet sensitivity to errors.}
    \begin{tabular}{|c|c|c|c|c|}
    \hline
        magnet type &  wiggler & sextupole & dipole & quadrupole\\
         \hline
         Translations & 10$\mu$m & 100$\mu$m & 500$\mu$m & 5$\mu$m\\
         Tilts & 0.5mrad & $>$25mrad &  0.5mrad & 1mrad\\
         \hline
    \end{tabular}
   
    \label{translationTolerances}
\end{table}

\subsection{Field Errors}

Tolerance studies determined the rms percent error in field strength permitted for most error seeds to reach the target dynamic aperture and are shown in Table \ref{MAerrorTable}.

\begin{table}[htb]
\caption{Field strength rms errors for at least $50\%$ seeds reaching target.}
\centering

\begin{tabular}{|c|c|c|}
\hline
Magnet Type & Parameter & Error \\ \hline
Dipole      & \textit{B} & 0.05\%             \\ \hline
Quadrupole  & $k_1$ & 0.2\%                    \\ \hline
Sextupole   & $k_2$ &  31\%                     \\ \hline
Wiggler     & $B_0$ & 0.03\%                     \\ \hline
\end{tabular}

\label{MAerrorTable}
\end{table}

\subsection{Orbit Correction}

Translations and tilts in wigglers and dipoles could largely be corrected by the two sets of BPMs and correctors on either side of each magnet. Quadrupole translation errors have the strictest tolerances of all magnet types, but as mentioned previously, this could be partially mitigated by a determination of which quadrupoles have the largest contribution to the reduction of DA. Sextupole tolerances are large compared to other magnet types and do not present any particular concerns.

\begin{figure}[htb]
    \centering
    \includegraphics[width = \linewidth]{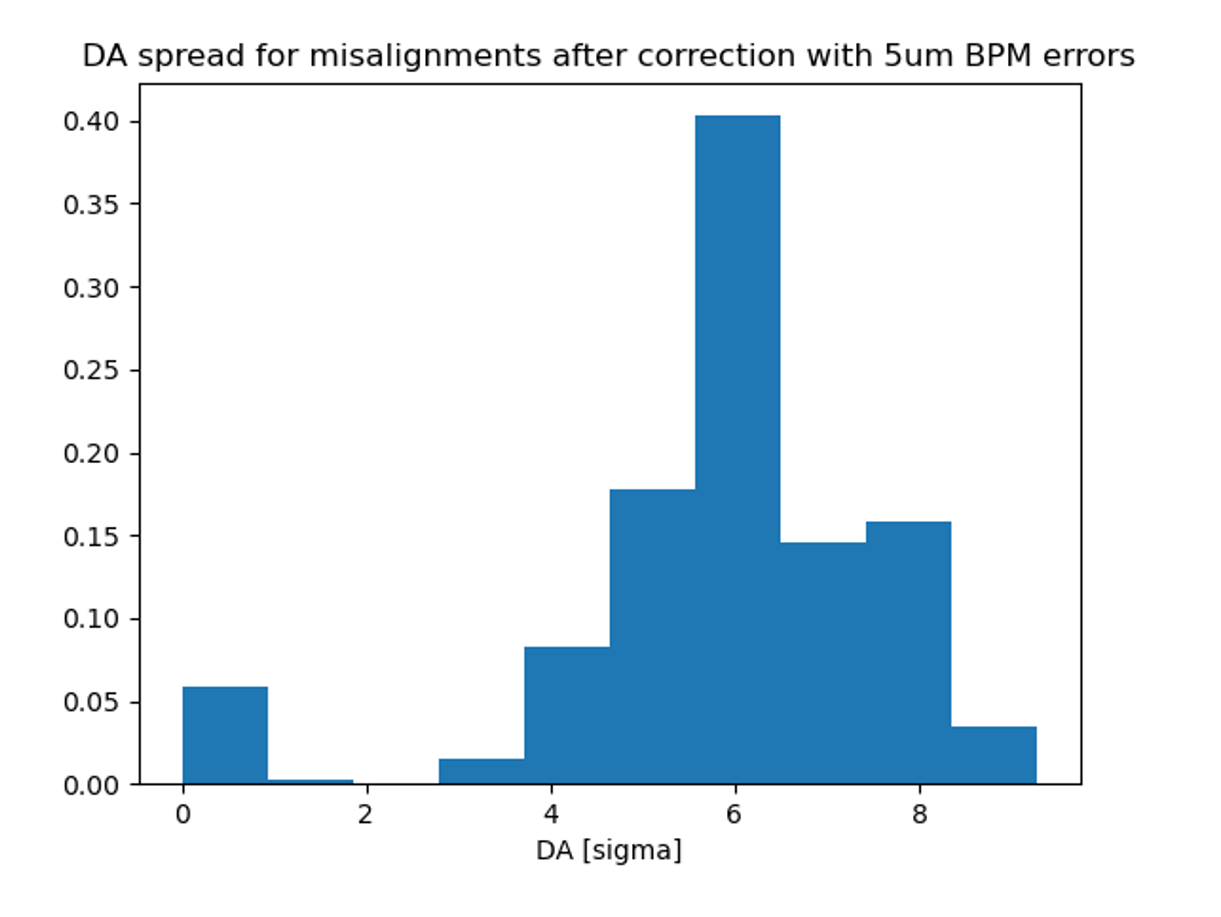}
    \caption{Dynamic aperture spread with misalignments from Table \ref{errors_with_correction} orbit correction and RMS BPM error of $5\mu$m.}
    \label{DA_with_misalignments_BPM_5um}
\end{figure}

\begin{figure}[htb]
    \centering
    \includegraphics[width = \linewidth]{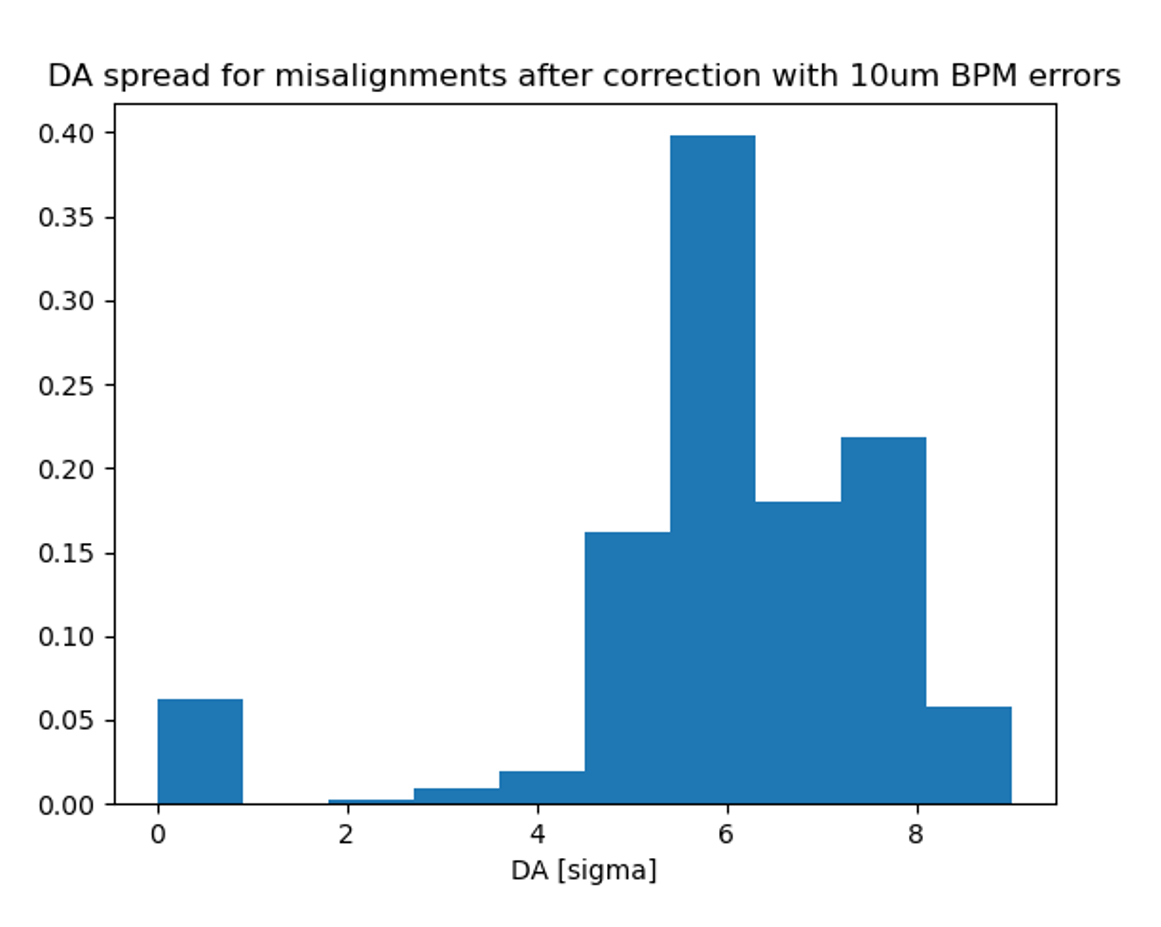}
    \caption{Dynamic aperture spread with misalignments from Table \ref{errors_with_correction}, orbit correction, and RMS BPM error of $10\mu$m.}
    \label{DA_with_misalignments_BPM_10um}
\end{figure}

The orbit correction scheme chosen for this study, while including translation errors for all magnet types, uses a BPM and a dual-plane corrector kicker at each quadrupole. An RMS BPM error also assumed during orbit correction. The splitting of errors between magnet types after correction are placed in table \ref{errors_with_correction}, and are chosen to keep $>$90$\%$ of runs above 5$\sigma$. This was achieved with BPM errors of 5$\mu$m or 10$\mu$m which are within the tolerances for modern BPM systems \cite{Singh2013NSLSIIBA} and shown in Figs \ref{DA_with_misalignments_BPM_5um} and \ref{DA_with_misalignments_BPM_10um}.

\begin{table}[htb]
\centering
\caption{Translation errors with orbit correction.}

\begin{tabular}{|c|c|c|}
\hline
Magnet Type & Translation Error \\ \hline
Dipole      & 100$\mu$m              \\ \hline
Quadrupole  & 50$\mu$m                     \\ \hline
Sextupole   & 100$\mu$m                    \\ \hline
Wiggler     & 50$\mu$m                      \\ \hline
BPM         & 5$\mu$m or 10$\mu$m    \\ \hline
\end{tabular}

\label{errors_with_correction}
\end{table}

\begin{figure}[htb]
    \centering
    \includegraphics[width = \linewidth]{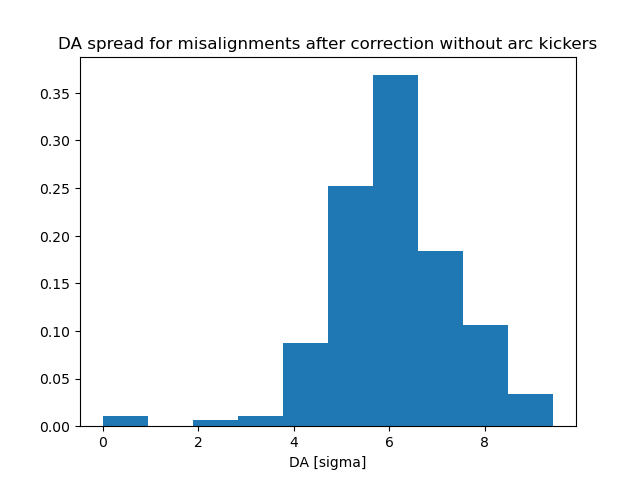}
    \caption{Dynamics aperture over error seeds from Table \ref{errors_with_correction} with BPMs and kickers at all quadrupoles except in arcs.}
    \label{no_kick_in_arcs}
\end{figure}

A potential concern in this orbit correction scheme is the placement of a kicker and BPM at each quadrupole is difficult to obtain in the arcs, where the separation between quadrupole and dipole magnets in the current lattice is 10cm. This correction was tested without BPMs and kickers in the arcs, leading to only a minor loss in dynamic aperture, with results compiled in Fig. \ref{no_kick_in_arcs} and show that a realistic correction scheme is available to correct DA with currently achievable magnet alignment and BPM accuracy.

\subsection{Emittance Increase from Errors}

\begin{figure}[htb]
    \centering
    \includegraphics[width=1\linewidth]{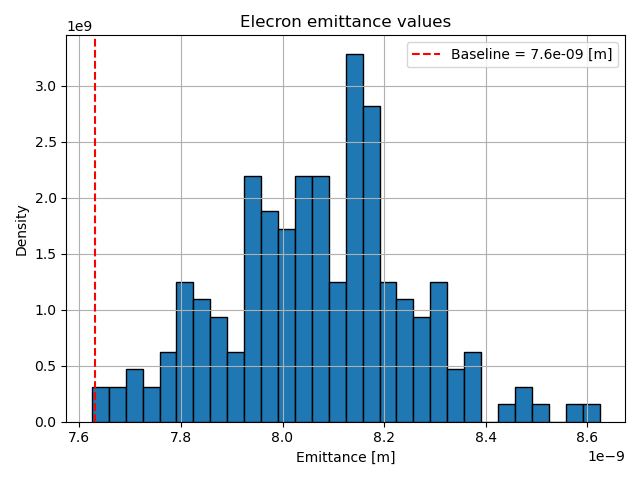}
    \caption{Emittance with IBS and BBS for 1000 error seeds using the misalignments from Table \ref{errors_with_correction} and orbit correction.}
    \label{emit_with_errors}
\end{figure}

As seen previously with the wiggler misalignments, keeping the orbit under control in the wigglers leads to small changes in the $\mathcal{H}_x$ function which can drive emittance increase. With the alignment errors and orbit correction applied from the previous section, the emittance was calculated for many error seeds using GETRAD8. This simulation included emittance increase from IBS and BBS from the proton beam using the latest cooling parameters\cite{seletskiy:napac2025-frad02}. The results of these runs are compiled in Fig. \ref{emit_with_errors} and show an emittance increase of $<10\%$ for almost all runs. This emittance increase is acceptable and keeps the emittance around the assumed 8nm used in cooling simulations.

\section{Space Charge Effects}

The REC uses a high bunch charge of 21nC in order to provide the required cooling rate for the proton beam. Initial estimates of the space charge tune shift were  -0.14 in each plane for the electrons in the storage ring \cite{seletskiy:napac2025-frad02}, giving concern about the effects on emittance growth and beam lifetime that this could lead to.

\subsection{Space Charge Model}

In order to investigate space charge in the REC, a suitable tracking method is required. A full 3D space charge model with frequent recalculation would be too slow for the tracking which would need to be at least one damping period, around 50000 turns. For electron beams that expect to stay roughly in a Gaussian distribution, a good approximation of the space charge kicks seen on particles in the beam is the Bassetti-Erskine model. While more often seen in the form for colliding beams, it can also be put into a convenient form for space charge:

\begin{multline}
        K_y+iK_x=\frac{r_eN}{\gamma^3\sigma_z}\text{exp}\left[\frac{-z^2}{2\sigma^2_z}\right]\sqrt{\frac{\sigma_x+\sigma_y}{\sigma_x-\sigma_y}}\left\{\text{erf}\left[\frac{x+iy}{\sqrt{2\left(\sigma_x^2-\sigma^2_y\right)}}\right]\right. \\
        \left.-\text{exp}\left[-\frac{x^2}{2\sigma_x^2}-\frac{y^2}{2\sigma_y^2}\right]\text{erf}\left[\frac{x\frac{\sigma_y}{\sigma_x}+iy\frac{\sigma_x}{\sigma_y}}{\sqrt{2\left(\sigma_x^2-\sigma^2_y\right)}}\right]\right\}
\end{multline}

This model will be the basis of tracking with space charge in the REC, although it is additionally useful to have an accurate estimate of the space charge tune shift expected in the lattice, as a sanity check and to determine the suitability of the model. The tune shift expected in a lattice can be shown to be

\begin{equation}
    \Delta Q_{x,y}=\frac{r_e}{4\pi}\frac{N}{\beta^2\gamma^3}\frac{F_{x,y}G_{x,y}}{B_f}\left<\frac{\beta_{x,y}}{\sigma_{x,y}\left(\sigma_x+\sigma_y\right)}\right>   
\end{equation}

\noindent where $B_f=\overline{\lambda}/\hat{\lambda}$ is the bunching factor, and $F_{x,y}$ and $G_{x,y}$ are factors related to the beam distribution and emittances and are tabulated for common examples \cite{Fedotov:techNote,Schindl:1999zr}. For the case of a Gaussian beam with equal transverse emittances, this goes to 

\begin{equation}
    \Delta Q_{x,y}=\frac{r_e}{4\pi}\frac{N}{\beta^2\gamma^3B_f}\left<\frac{\beta_{x,y}}{\sigma_{x,y}\left(\sigma_x+\sigma_y\right)}\right>
    \label{tuneShift}
\end{equation}

\noindent where for a bunch with a longitudinally Gaussian profile, $B_f=\sqrt{2\pi}\sigma_z/L$, with $L$ being the total ring length. This can then be applied to the REC lattice and bunch charge, yielding an expected tune shift of -0.079/-0.115 in the horizontal and vertical directions respectively. This tune shift, while less than the initially assumed $-0.14$ in each plane during the early design stages, is large and must be studied to ensure adverse effects are identified and corrected.

\subsection{Tracking with Space Charge}

\begin{figure}[ht]
    \centering
    \includegraphics[width=0.9\linewidth]{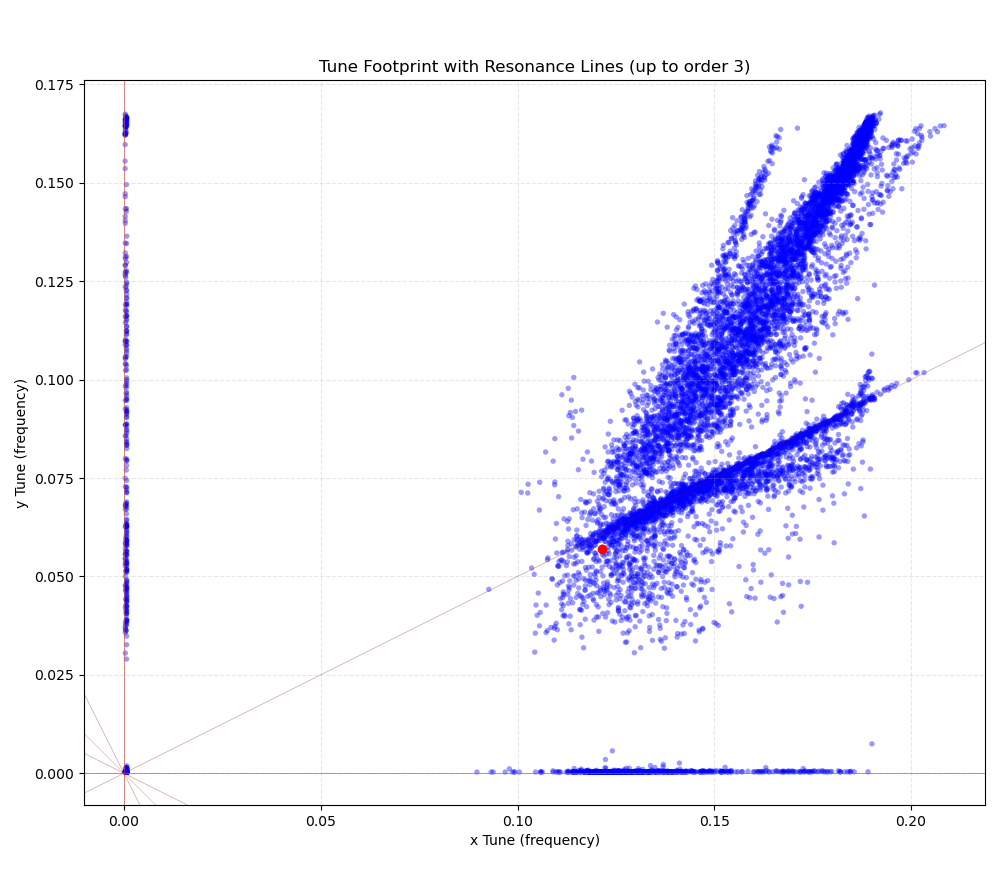}
    \caption{Tune footprint due to space charge. Estimated tune shift of the bunch core shown in red.}
    \label{tuneFootprint}
\end{figure}

In order to measure the space charge tune shift in the bunch, a 5000 particle bunch was tracked over 2000 turns, with the tune of the particles determined by a NAFF algorithm of the final 1000 turns. The space charge model was the before mentioned Bassetti-Esrkine with the assumption of the constant emittance. The result of this tracking was the tune footprint in Fig \ref{tuneFootprint}, where it can be seen that the highest shifted particles appear near the estimated tune shift. It can be seen that the maximum tune shift sits just past a strong sextupole resonance, with many particles sitting along the resonance line.

\begin{figure}[ht]
    \centering
    \includegraphics[width=0.9\linewidth]{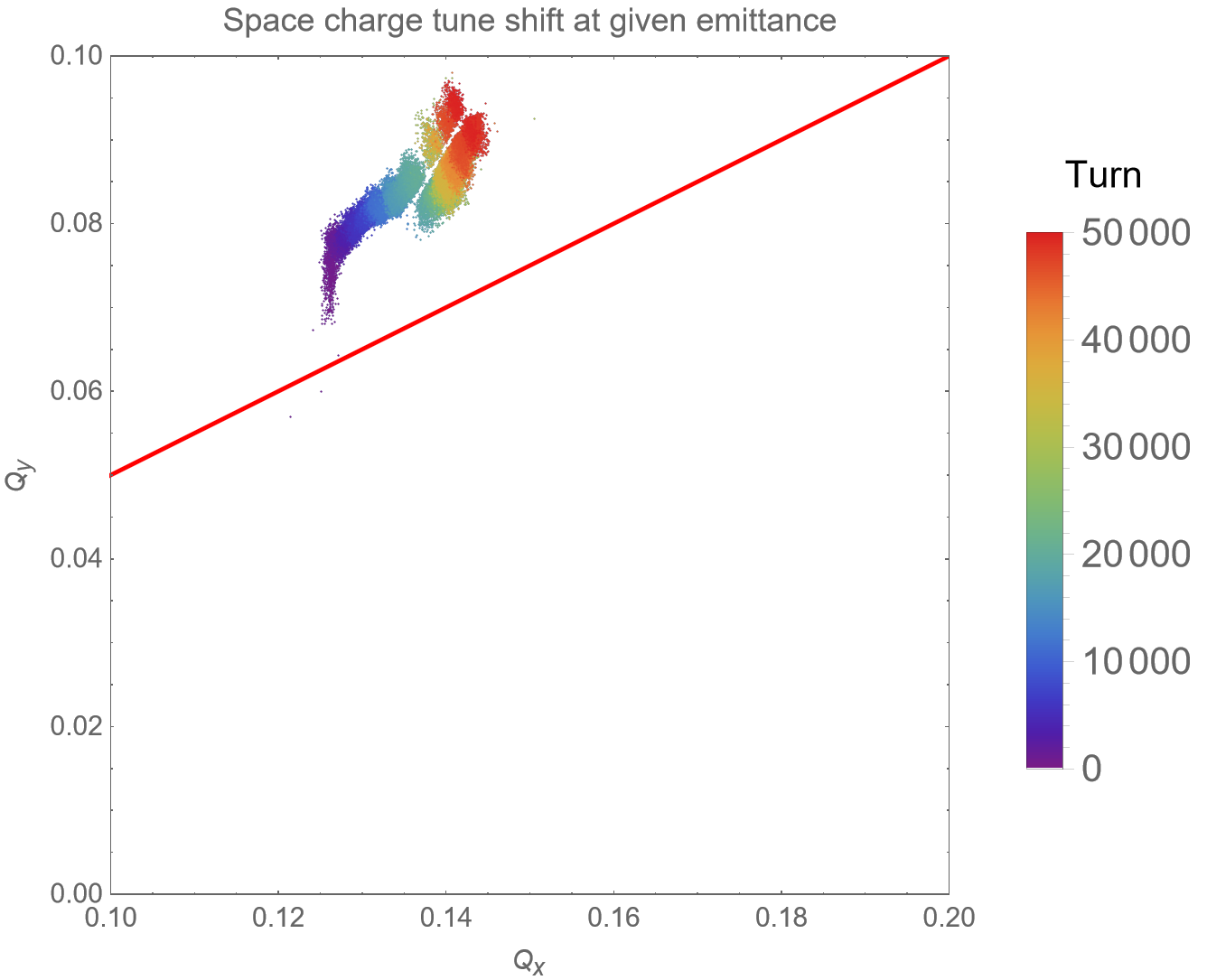}
    \caption{Location of the space charge tune shift as the emittance of the bunch increases over 50000 turns. Note the resonance jumping starting at turn 20,000.}
    \label{ChangeInTuneShift}
\end{figure}

While using the assumption of a constant emittance is appropriate for determining the initial tune shift, in order to study long term effects, any change in the size of the beam should be accounted for. This was done by updating the emittance used in the Bassetti-Erskine model every 1000 turns, which was small when compared to multiple damping periods. This changes the space charge kick overtime, and, for a beam with presumably increasing emittance, this will lead to a weakening of the space charge tune shift. This can be shown in Fig. \ref{ChangeInTuneShift} where the core tune shift is plotted over the run. This tune shift can also be seen to start crossing a resonance around turn 20,000

\begin{figure}[ht]
    \centering
    \includegraphics[width=1\linewidth]{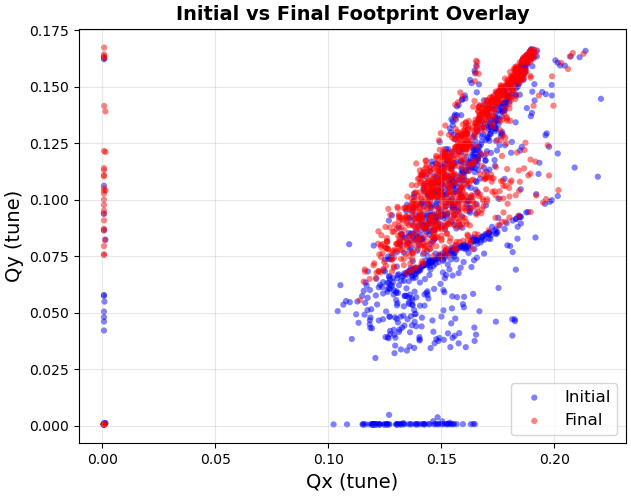}
    \caption{Tune Footprint decreasing in size as the space charge tune shift decreases from emittance increase.}
    \label{FootPrintRecalcComp}
\end{figure}

The effect this has on the full bunch is shown in Fig. \ref{FootPrintRecalcComp}, where it can be seen that the tune footprint begins to shrink, with particles beginning to leave the region of the sextupole resonance. This resonance crossing did not cause particle loss in excess of the static space charge model, showing that dynamic changes in the space charge distribution do not cause additional beam instability noticeable within two damping periods. 

\subsection{Emittance Growth}

\begin{figure}[htbp]
    \centering
        % Top subfigure
    \begin{subfigure}{0.9\columnwidth}
        \centering
        \includegraphics[width=\linewidth]{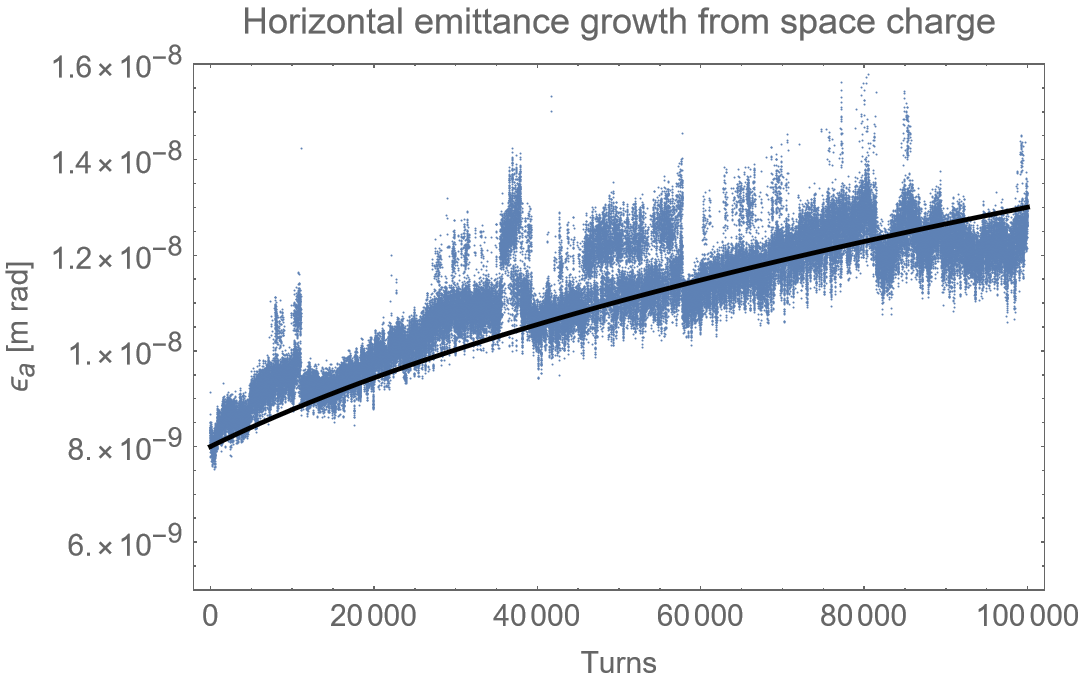}
        \caption{Horizontal}
        \label{ExIncrease}
    \end{subfigure}

    % Bottom subfigure
    \begin{subfigure}{0.9\columnwidth}
        \centering
        \includegraphics[width=\linewidth]{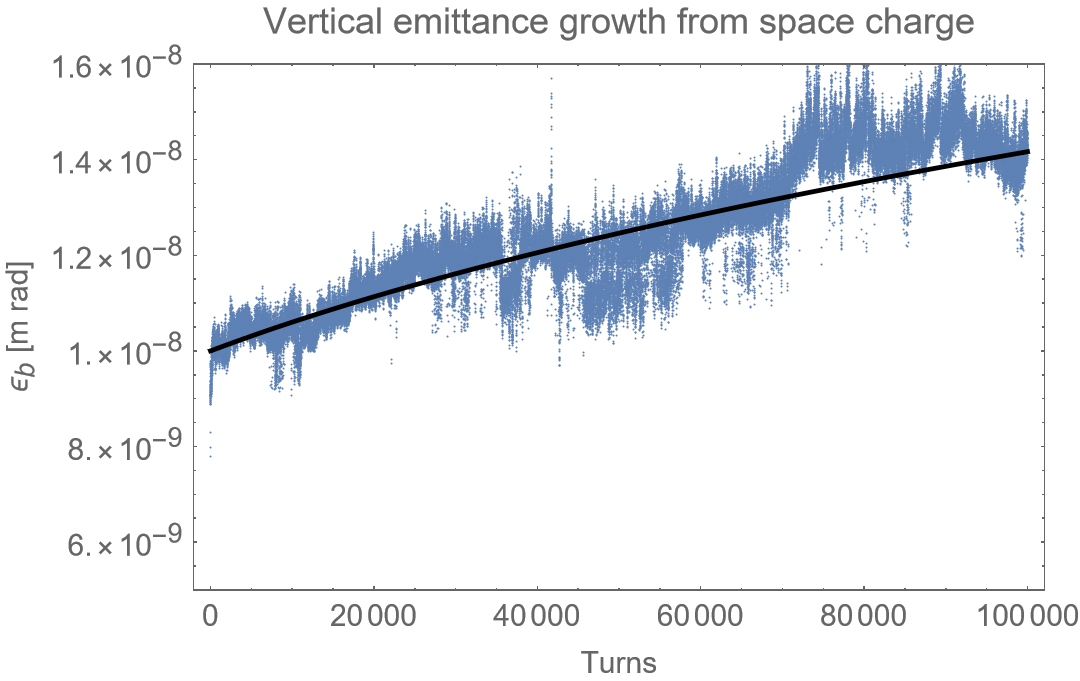}
        \caption{Vertical}
        \label{EyIncrease}
    \end{subfigure}
    \caption{Emittance increase from space charge with recalculation every 1000 turns.}
    \label{space charge emittance increase}
\end{figure}

One potential challenge with space charge is the increase in emittance that the beam can see over time. As mentioned previously, this change results in a decrease of the space charge tune shift so we expect this growth to decrease over time. An approximation for the scaling can be found by assuming emittance growth of tune diffusion with a scaling of $\Delta Q^2$. It is then seen from eq. \ref{tuneShift} that the growth rate would be

\begin{equation}
    R_{SC x,y}=\frac{A_{SCx,y}}{\epsilon_{x,y}\left(\sqrt{\epsilon_x}+\sqrt{\epsilon_y}\right)}
\end{equation}

\noindent where $A_{SCx,y}$ is a scaling factor which will be taken from tracking. This tracking was done for $10^5$ turns with the same conditions as before, leading to Figs. \ref{ExIncrease} and \ref{EyIncrease}. both of these saw increase over the period on the order of 50\%. This tracking neglects the effects of other sources of emittance change like IBS, BBS, and radiation, so to find a new estimate of the equilibrium emittance, these effects must be balanced. 

As the emittance scaling of all of these effects is known, it is possible to arrive at an approximation of the equilibrium by finding the value such that the sum of the rates is zero

\begin{equation}
    R_{SC x,y}+R_{IBS}+R_{BBS}+R_{RADx,y}+R_0=0
\end{equation}

\noindent where $R_0$ is the stochastic part of the radiation and each other terms are

\begin{equation}
    R_{IBS}=\frac{A_{IBS}}{\epsilon_x\epsilon_y}
\end{equation}

\begin{equation}
    R_{BBS}=\frac{A_{BBS}}{\epsilon_x\epsilon_y}
\end{equation}

\begin{equation}
    R_{RADx,y}=A_{RADx,y}\epsilon_{x,y}
\end{equation}

\noindent where the $A$'s are scaling factors taken from tracking. This gives an equilibrium emittance of 8.16 and 7.88 [rad nm] where less than 1\% is contributed from space charge, showing that this emittance growth is substantially less than from IBS and BBS, meaning previous emittance estimates realistically support the cooling rates expected in the REC design and does not pose an additional concern.

\subsection{Image Charge and Current}

In addition to the direct space charge of the previous sections, indirect effects such as image charge or image currents also must be studied. Tune shifts from image charge can largely be reduced through the use of circular beam pipes, however image currents induced in the wiggler pole faces are not as easily reduced since wigglers account for almost a fifth of the length of the ring. The image charge and current tune shifts are an additional term in eq. \ref{tune_shift}

\begin{multline}
        \Delta Q_{x,y}=-\frac{Nr_0}{4\pi \beta^2\gamma}\left(\left\langle\frac{\beta_{x,y}}{\sigma_{x,y}\left(\sigma_x+\sigma_y\right)}\right\rangle\frac{\varepsilon^{x,y}_0}{\gamma^2B_f}\right. \\ + \left.\left\langle\beta_{x,y}\right\rangle\frac{\varepsilon^{x,y}_1}{h^2B_f}+\left\langle\beta_{x,y}\right\rangle\beta^2\frac{\varepsilon^{x,y}_2}{g^2}\right)
        \label{tune_shift}
\end{multline}

\noindent where $h$ is the beam pipe height, $g$ is the pole gap, and the $\varepsilon$'s are the Laslett coefficients. With the optics in the wigglers as shown previously in fig. \ref{WigglerSectionOptics} it can be seen that the tune shift from image currents is expected to be significantly larger in the vertical direction.

\begin{figure}[htbp]
    \centering
        % Top subfigure
    \begin{subfigure}{0.9\columnwidth}
        \centering
        \includegraphics[width=\linewidth]{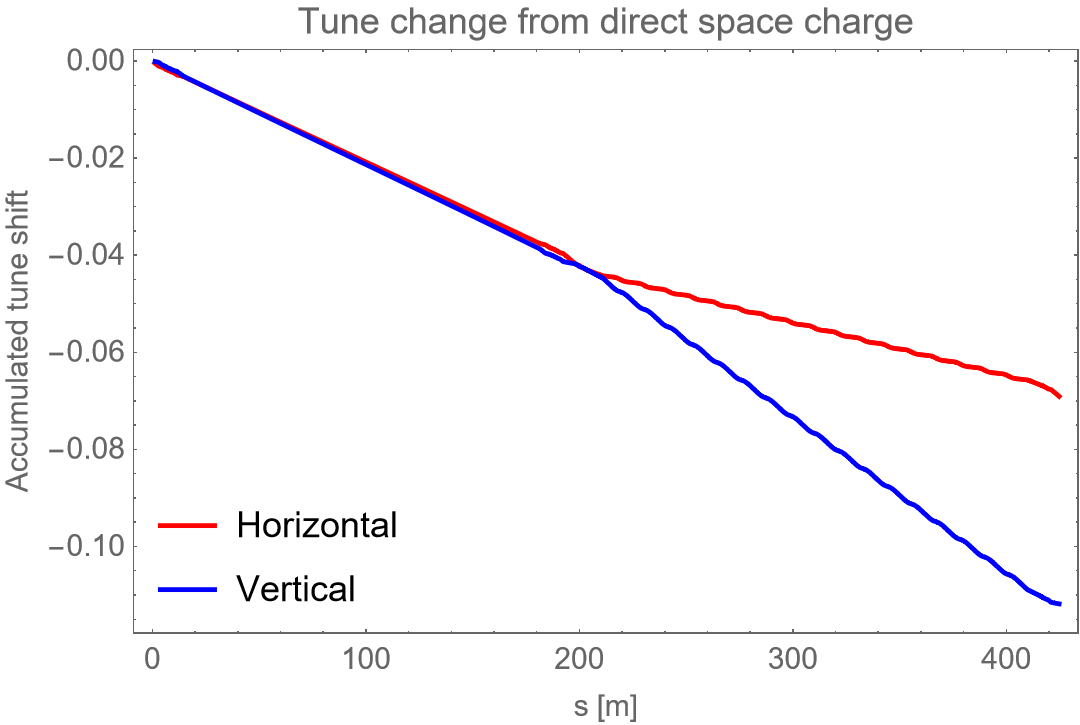}
        \caption{Tune shift from direct space charge}
        \label{direct space charge}
    \end{subfigure}
    \begin{subfigure}{0.9\columnwidth}
        \centering
        \includegraphics[width=\linewidth]{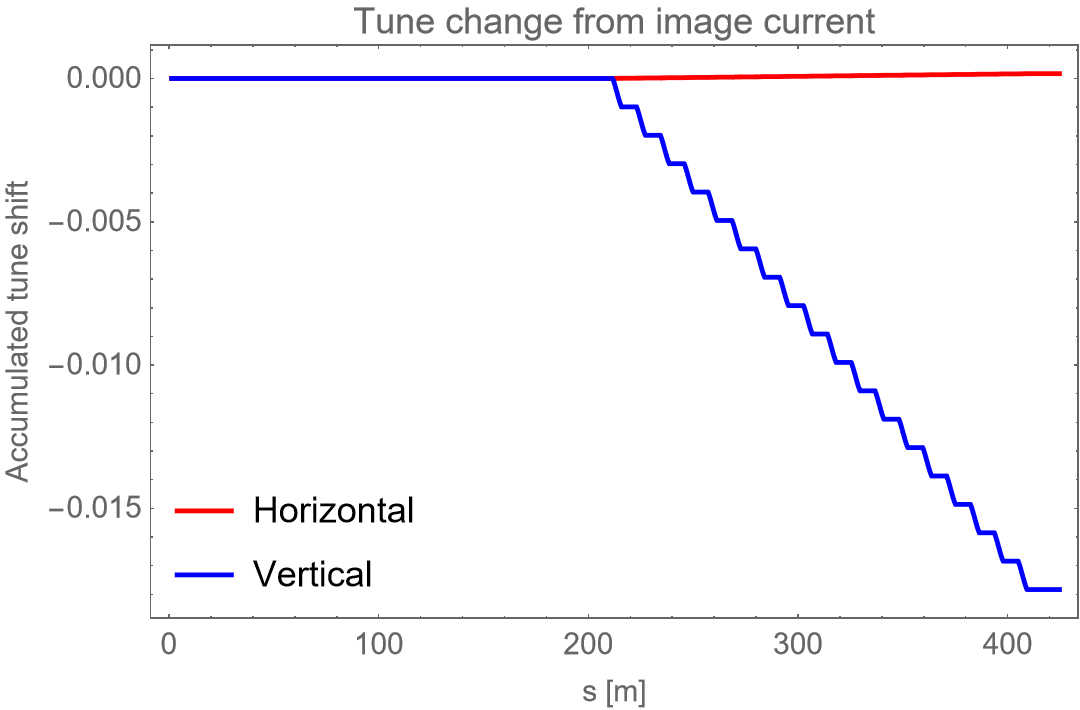}
        \caption{Tune shift from image currents}
        \label{image current accumulation}
    \end{subfigure}
    \caption{Space charge tune shift integrated over the ring from the two largest sources.}
    \label{image current}
\end{figure}

When the $\beta$-function of the wiggler section is used with a pole gap of 1cm, the resulting tune shift is $1.7\times10^{-4}$ in the horizontal and $-1.8\times10^{-2}$ in the vertical direction. This tune shift is largely negligible in the horizontal direction, but the shift in the vertical direction could cause noticeable but minor changes in dynamics when compared to the direct space charge tune shift. When integrated through the ring this yields fig. \ref{image current}. As this tune shift is about half of the vertical phase advance in the wigglers, wiggler focusing can be adjusted if needed to largely counteract this focusing, which, unlike direct space charge which is non-linear for particles outside of 1$\sigma$, is linear for particles of small amplitude in relation to the pole gap of 1cm.

\section{Conclusion}

A nonlinear dynamics and optimization study has been presented for the EIC Ring Electron Cooler (REC) operating at 150 MeV for cooling of 275 GeV protons. The REC lattice combines a long cooling section, strong radiation damping wigglers, and compact geometry while maintaining acceptable dynamic aperture and tolerance to collective effects.

A primary challenge of the design is the strong nonlinear dynamics introduced by the focusing wigglers required to suppress emittance growth from  intra-beam scattering and beam-beam scattering. A sextupole-like focusing wiggler field configuration was shown to substantially reduce chromatic effects, as compared to simple quadrupole-like focusing, where the singularity in the chromaticity of the wiggler with superimposed quadrupole caused extreme chromaticity in the vertical plane. Switching fields to the sextupole-like focusing was advantageous to increase the DA of the REC, although novel applications of such a field may exist. Combined optimization of the sextupole configuration, octupole compensation, and phase advance through the wiggler blocks yielded transverse aperture exceeding the operational $5\sigma$ target and yielded momentum acceptance approaching the $5\sigma_{p_z}$ target.

Sensitivity studies including realistic alignment and field errors demonstrated that the optimized lattice remains operable with achievable magnet tolerances and BPM resolutions using a practical orbit correction scheme. Tracking studies including direct space charge showed tune shifts consistent with analytic expectations. They indicated that the resulting emittance growth remains small compared with contributions from intra-beam scattering and beam-beam scattering. Indirect effects from image currents in the wigglers were also shown to produce only moderate additional perturbations to the optics.

These results demonstrate the feasibility of ring-based electron cooling for the EIC and establish a viable lattice framework for future optimization. Further work will focus on increasing the momentum acceptance margin, improving robustness against machine errors, and incorporating additional collective effects and self-consistent cooling dynamics into long-term tracking studies.

\bibliography{apssamp}% Produces the bibliography via BibTeX.

\end{document}